\documentclass{aastex}
\usepackage{emulateapj5}
\usepackage{onecolfloat5}
\usepackage{natbib}
\bibpunct{(}{)}{;}{a}{,}{,}

\def\be{\begin{equation}}
\def\ee{\end{equation}}
\def\bea{\begin{eqnarray}}
\def\eea{\end{eqnarray}}
\def\la{\mathrel{\mathpalette\fun <}}

\def\fun#1#2{\lower3.6pt\vbox{\baselineskip0pt\lineskip.9pt
  \ialign{$\mathsurround=0pt#1\hfil##\hfil$\crcr#2\crcr\sim\crcr}}}

\input epsf
\def\plotone#1{\centering \leavevmode
\epsfxsize= 1.0\columnwidth \epsfbox{#1}}

\def\gsim{\;\rlap{\lower 2.5pt
 \hbox{$\sim$}}\raise 1.5pt\hbox{$>$}\;}
\def\lsim{\;\rlap{\lower 2.5pt
   \hbox{$\sim$}}\raise 1.5pt\hbox{$<$}\;}

\newcommand{\beq}{\begin{equation}}
\newcommand{\eeq}{\end{equation}}
\def\myputfigure#1#2#3#4#5%
{\hskip0.03\textwidth\vskip#5pt
\makebox[0pt]{\hskip#2in
\includegraphics[width=#3\textwidth]{#1}}\vskip#4pt\hfill}


\begin{document}
\bibliographystyle{apj}
\twocolumn[
\submitted{Submitted to ApJ}

\title{Probing the Reionization History of the Universe using the 
Cosmic Microwave Background Polarization}

\author{Manoj\ Kaplinghat\altaffilmark{1}, 
Mike\ Chu\altaffilmark{1},
Zolt\'an\ Haiman\altaffilmark{2,3,5}, 
Gilbert\ P.\ Holder\altaffilmark{4}, 
Lloyd\ Knox\altaffilmark{1} and 
Constantinos\ Skordis\altaffilmark{1}} 

\vspace{\baselineskip}

\begin{abstract}
The recent discovery of a Gunn--Peterson (GP) trough in the spectrum of
the redshift 6.28 SDSS quasar has raised the tantalizing possibility
that we have detected the reionization of the universe.  However, a
neutral fraction (of hydrogen) as small as 0.1\% is sufficient to cause
the GP trough, hence its detection alone cannot rule out
reionization at a much earlier epoch. The Cosmic Microwave Background
(CMB) polarization anisotropy offers an alternative way to explore the
dark age of the universe. We  show that for most models constrained by
the current CMB data and by the discovery of a GP trough
(showing that reionization occurred at $z > 6.3$), MAP can detect the
reionization signature in the polarization power spectrum. The expected
1-$\sigma$ error on the measurement of the electron optical depth is
around 0.03 with a weak dependence on the value of that optical
depth. Such a constraint on the optical depth will allow MAP to
achieve a 1-$\sigma$ error on the amplitude of the primordial power
spectrum of 6\%. MAP with two years (Planck with one year) of
observation can  distinguish a model with 50\% (6 \%) partial
ionization between redshifts of 6.3 and 20 from a model in which
hydrogen was completely neutral at redshifts greater than 6.3.  Planck
will be able to distinguish between different reionization histories
even when they imply the same optical depth to electron scattering for
the CMB photons.    

\end{abstract}
 ]

\altaffiltext{1}{Department of Physics, 1 Shields Avenue,
University of California, Davis, California 95616, USA}
\altaffiltext{2}{Princeton University Observatory, Princeton, NJ 08544,
USA}
\altaffiltext{3}{Department of Astronomy, Columbia University, 550 West
120th Street, New York, NY 10027, USA} 
\altaffiltext{4}{Institute for Advanced Study, School of Natural
Sciences, Olden Lane, Princeton, NJ 08540, USA}
\altaffiltext{5}{Hubble Fellow}

\section{Introduction}
\label{sec:introduction}

How and when the intergalactic medium (IGM) was reionized is one of the
long outstanding questions in cosmology, likely holding many clues about
the nature of the first generation of light sources and the end of the 
cosmological ``Dark Age'' \citep[see][for a review of our current
understanding]{barkana01}.  The lack of strong HI absorption (the
``Gunn--Peterson '', GP, trough) in the spectra of high
redshift quasars has revealed that the intergalactic medium (IGM) is
highly ionized between redshifts $0\lsim z \lsim 6$.  On the other hand,
the lack of a strong damping by electron scattering of the first
acoustic peak in the temperature anisotropy of the cosmic microwave
background (CMB) radiation has shown that the universe was neutral
between the redshifts $30\lsim z \lsim 10^3$.  Together these two sets
of data imply that most hydrogen atoms in the universe were reionized
during the redshift interval $6\lsim z \lsim 30$. 

The recent discovery by \cite{becker01} of the bright quasar SDSS
1030+0524 in the Sloan Digital Sky Survey (SDSS) at redshift $z=6.28$
has, for the first time, revealed a full GP trough, i.e., a spectrum
consistent with no flux at a substantial stretch of wavelength shortward
of $(1+z)\lambda_\alpha=8850$\AA.  This discovery has raised the 
tantalizing possibility that we are detecting reionization occurring near
redshift $z\sim 6.3$.  The lack of any detectable flux indeed implies a
strong lower limit $x_{\rm H}\gsim 0.01$ on the mean mass--weighted
neutral fraction of the IGM at $z\sim 6$ \citep{fan02, pentericci02}.
On the other hand, because of the large opacities of the hydrogen
Lyman series, it is considerably difficult to push this method much
further. In particular, in order to prove that we are probing into the
neutral epoch, one would like to directly infer $x_{\rm H}\approx 1$.
While in principle it is possible to infer this from the lack of any
flux in a high enough S/N spectrum, in practice the required
integration times are implausibly long. 

In this paper, we discuss an alternative way of probing deeper into the
dark ages. CMB polarization anisotropy at large angles is very
sensitive to the optical depth to electron scattering for the CMB
photons \citep[and references
therein]{basko80,hogan82,zaldarriaga97b,haimanknox99}. 
In a model with no reionization, the polarization signal at large angles
is negligible. However, CMB photons scattering in a reionized medium
greatly enhance the polarization signal  -- making it very likely that 
such a signal (based on present CMB temperature anisotropy data and the 
GP trough) will be detected by the ongoing CMB satellite experiment
MAP. We also show that Planck (and for large optical depths, MAP), will
have the power to discriminate between different reionization histories
even when they lead to the same optical depth. 

The results from CMB polarization experiments will deepen our
understanding of the physics of reionization. They will lead to better
constraints on the models of reionization and, thereby, tell us much
more about the first sources of light, and indeed, the first structures
to form. 

\section{Probes of the dark age}

Very little is known about the first sources and the reionization of
the universe. Given the existence of the GP trough in the spectrum of
the z=6.28 quasar, it is possible to argue on the basis of theoretical 
models and numerical simulations that reionization indeed occurred
close to $z=6.3$.  The key ingredient in such arguments is that neutral
fractions inferred from a sample of high redshift quasars show a rapid
rise with redshift over the range $5.5\lsim z \lsim 6$
\citep{songaila02}.  
Semi--analytical methods \citep[for example]{haiman98}, as well as
numerical simulations \citep[for example]{gnedin98, gnedin99} suggest
that the initial  phase of reionization proceeds very rapidly. During
the ``overlap'' phase, when the individual HII regions of the ionizing
sources percolate, the mean neutral fraction drops from unity by several
orders of magnitude in a small fraction of the Hubble time.  The reason
for this rapidity is the fast timescale for the formation rate of the 
ionizing sources, which correspond to the high-$\sigma$ peaks of the
original density field.  More recently, simulations of cosmological
reionization \citep{cen02,gnedin01,fan02}, geared specifically to
interpret the spectrum of SDSS 1030+0524, argued that the IGM is likely
neutral at $z\gsim 6.5$. 

While there may be a theoretical bias for reionization occurring close to 
$z=6.3$, it is conceivable for reionization to last over a considerably
longer redshift interval.  This could be the case if the formation rate
of the ionizing sources does not parallel the collapse of high-$\sigma$
peaks, but is more gradual.  Complete reionization could also be delayed
for $\Delta z \gsim 1$ due to screening by minihalos with virial
temperatures less than $10^4 \ K$ \citep{haiman01,barkana02}. In
addition, the decrease of the mean neutral fraction would be more
gradual if the ionizing sources had a hard spectrum.  Reionization by
X--rays was considered recently by \citet{oh01} and
\citet{venkatesan01}. In contrast to a picture in which discrete HII
regions eventually overlap, in this case, the IGM is ionized uniformly
and gradually throughout space. 

An observational probe of the redshift--history of reionization would be
invaluable in constraining such scenarios, and to securely establish
when the Dark Age ended.  Although detections of the hydrogen GP trough
can not provide this type of constraint, a possibility may be to use
corresponding absorption troughs caused by heavy elements in the high
redshift IGM.  Recently, \citet{oh02} showed that if the IGM is
uniformly enriched by metals to a level of $Z=10^{-2}-10^{-3}~{\rm
Z_\odot}$, then absorption by resonant lines of OI or SiII may be
detectable.  The success of this method depends on the presence of
oxygen and silicon at these abundance levels, and also requires that
metal enrichment precede hydrogen reionization. 

An alternative method to probe the reionization history is to utilize
the systematic changes in the profiles of Lyman alpha emission lines
towards higher redshift. As argued in \citet{haiman02a}, the increasing
hydrogen IGM opacity towards higher redshift would make the emission
lines appear systematically more asymmetric, and the apparent
line-center systematically shifted towards longer wavelengths, as
absorption in the IGM becomes increasingly more important, and
eliminates the blue side of the line.  Because of the  intrinsically
noisy Ly alpha line shapes, this method will require a survey that
delivers a large sample of Ly alpha emitting galaxies \citep{rhoads01}.

Finally, it has been suggested that future radio telescopes could
observe 21 cm emission or absorption from neutral hydrogen at the
time of reionization \citep[e.g.,][]{hogan79,tozzi00}. 
This would provide a direct measure of
the physical state of the neutral hydrogen and its evolution through
the time of reionization. The detectability of the signal depends
sensitively on the details of reionization and will require isolation
of a signal that is much smaller than foreground contaminants.
Recently, \citet{carilli02} also considered the radio equivalent of the
GP trough. Unlike the Ly$\alpha$ case, the mean absorbtion by the
neutral medium is about 1\% at the redshifted 21 cm. While
\citet{carilli02} considered the 21 cm absorbtion from large-scale
structure using simulations, \citet{furlanetto02} and \citet{iliev02}
have used semi-analytic methods to look at the observable features (in
21 cm) of minihalos and protogalactic disks. These studies argue that
the 21 cm observations would yield robust information about the
thermal history of collapsed structures and the ionizing background,
sufficiently bright radio-loud quasars exist at $z_e > 6.3$. 
 
As mentioned in the introduction, an alternative method to probe
reionization is via the CMB anisotropy.  The advantage of this method
is that it probes the presence of free electrons, and can therefore
detect $x_{\rm H}=0.1$ and $x_{\rm H}=10^{-3}$ with nearly equal
sensitivity.  Physically, the CMB and the GP trough therefore probe two
different stages of reionization: the CMB is sensitive to the initial
phase when $x_{\rm H}$ first decreases below unity, and free electrons
appear, say, at redshift $z_e$ (see Figure \ref{fig:2step}).  On the
other hand, the (hydrogen) GP trough is sensitive to the end phase,
or, going backward in cosmic time, when neutral hydrogen atoms first
appear; say, $z_{\rm H}$.  In currently popular theories, these two
phases coincide to $\lsim 10$ percent of the Hubble time,
i.e. $z_e\approx z_{\rm H}$.  However, as argued above, one can
conceive alternative theories in which the two phases are separated by
a large redshift interval, and  $z_e\gg z_{\rm H}$. 

Our goal in this paper is to quantify the power of CMB anisotropy to
tell the difference between $z_e\gg z_{\rm H}$ and $z_e\approx z_{\rm 
H}\approx 6.3$, and more broadly, {\it whether there was substantial
activity in the universe prior to z=6.3}.  A similar question was
addressed recently by Gnedin \& Shandarin (2002), who consider the
effects of reionization on maps of the temperature anisotropy.  Our
work is complementary, in that we focus on signatures of an extended
period of partial reionization on the polarization anisotropy. There
has also been some recent work on the topic of extended reionization
by \citet{bruscoli02}. Their results stressed that although CMB
experiments will determine $\tau$, the redshift of reionization will
remain uncertain since that depends on the reionization
history. We point out here that in fact there is enough information in
CMB polarization to enable a detailed study of the reionization
history.    

In addition to probing high--redshift structure formation, our
calculations have importance for fundamental cosmology.  In particular,
the presence of free electrons damps the primary anisotropy, and
results in a strong degeneracy between the amplitude of primordial
fluctuations ($A$) and the electron scattering optical depth
($\tau$). The quantity $A$ is one of the most important parameters in
inflationary cosmology, as it directly probes the inflaton
potential. We show that the knowledge we gain about $\tau$ from 
MAP data will fix $A$ with a precision of about $5\%$.

The $\tau-A$ degeneracy and the effect of CMB polarization on it
has been studied before
\citep{zaldarriaga97,wang99,eisenstein99,tegmark00,prunet00,venkatesan02},
using Fisher matrix methods.
The resulting measures of $\tau$ and $A$ have been shown to be powerful
constraints on models of reionization \citep{venkatesan00,venkatesan02},
and it has also been shown that introducing some reasonable constraints
on models of reionization can allow improved estimates of
cosmological parameters.
In previous work, the epoch of reionization has been treated as a
sharp transition at a single redshift, leading to all information
on reionization from CMB measurements encoded by a single number,
$\tau$. We show below that there is significantly more information
in the large angle polarization anisotropy measurements. This will
eventually allow more sophisticated tests of models of reionization.

\section{Reionization signatures in the CMB}

Reionization affects the CMB anisotropy in a simple manner. On large
angular scales, reionization has no effect, while on
smaller scales it damps the anisotropy by a factor of
$\exp(-2\tau)$. The division into large and small scales is dictated by
the angle subtended by the horizon at reionization. The multipole moment
this angular scale maps onto is given by  $l_r = D_A(z(\eta_r))/\eta_r$,
where $\eta_r$ is the visibility function weighted conformal time
\citep{hu97} and $D_A(z)$ is the angular diameter distance to redshift
$z$. The above assumes that the photons scattering off of the
reionized electrons do not pick up additional anisotropy. While not
strictly correct, this approximation is good for $\tau \la 0.3$. The
electron scattering optical depth is given by 
\begin{eqnarray}
\nonumber
\tau & = & \int_0^{z_e}dz \sigma_{\rm T} n_e(z) c\frac{dt}{dz}\\
\tau & = & 0.038 \omega_b  h/ \omega_m \left[\left(\Omega_\Lambda+
\Omega_m\left(1+z_e\right)^3\right)^{1/2}-1\right],
\label{eq:taue}
\end{eqnarray}
where $\sigma_{\rm T}$ is the Thompson cross section, $n_e$ is the
electron abundance, and $\omega_b\equiv\Omega_b h^2$ and
$\omega_m\equiv\Omega_m h^2$ are the baryon and matter densities
(respectively) scaled to the critical density today.
Eq.~(\ref{eq:taue}) is correct for flat $w=-1$ models \citep{hu97}, and
we have assumed complete ionization of H and $^4$He to H$^+$ and
$^4$He$^+$ respectively out to a redshift of $z_e$. 

The effects of reionization on CMB polarization anisotropy are similar
to that on the temperature anisotropy. At $l \gg l_r$, the polarization
angular power spectra are suppressed by $\exp(-2\tau)$. However, in
contrast to the temperature power spectrum, the effect at $l \la l_r$ is
quite dramatic (see Figure \ref{fig:spectra}). Note that there is no
analog of the Sachs-Wolfe (or the integrated Sachs-Wolfe) effect for 
polarization. All of the polarization signal is generated at the last
scattering surface at $z\sim 1,000$ in the absence of reionization. Thus
without reionization, the polarization power at small multipoles is
negligible. In the reionized universe, the rescattering of
photons in the presence of a large quadrupole temperature anisotropy
results in a broad peak in polarization power at low $l$. For a detailed
discussion of this ``reionization bump'', we refer the reader to 
\citet{zaldarriaga97b}. Here we provide an overview. 

We first note that since  most of the optical depth to reionization is
generated prior to the onset of curvature or dark energy domination,
the reionization bump depends on curvature and dark energy only through
their effect on $D_A$. The polarization anisotropy is sourced by the
total quadrupole anisotropy -- which for practical purposes is just the
temperature quadrupole. The relevant quantity is the quadrupole
at the reionization surface. The quadrupole produced at the last
scattering surface (as the tight coupling breaks down) is small; however
free-streaming after last scattering transfers power from the monopole
to higher multipoles thus producing an appreciable quadrupole during
reionization. The modes contributing to the reionization
signal are $k \sim 2/\eta_r$ which become sub-horizon well after
recombination. Thus the quadrupole on these scales does not depend on
the baryon density. It does depend on the matter density when 
$\omega_m \la 0.05$ since then there's significant radiation around when
the relevant modes enter the horizon and the integrated Sachs-Wolfe
effect will change the quadrupole. For the most part, the
reionization bump is only sensitive to the electron optical depth and
the amplitude of primordial potential fluctuations. 

The arguments outlined above allow us to obtain the low-$l$ E mode
polarization anisotropy spectrum and its cross term with the temperature
spectrum by simply scaling from a fiducial model. The temperature 
auto-correlation, polarization auto-correlation and the
temperature-polarization cross-correlation spectra are denoted by 
$C_{T l}$, $C_{E l}$ and  $C_{C l}$ respectively. The $E$ subscript
refers to the E (electric or scalar) mode and the corresponding
pseudo-scalar B mode is absent from scalar perturbation generated
polarization anisotropy \citep{kamionkowski97,zaldarriaga97c}.  In terms 
of ${\cal C}_{X l} \equiv l(l+1)C_{X l}/2\pi$ (where $X$ stands for  
$E$,$T$ or $C$), the scaling from the fiducial model (denoted by
``$*$'') is given by 
\citep{kaplinghat02}:  
\begin{eqnarray}
\label{eq:pol}
{\cal C}_{El'} & = & {\cal C}_{E_* l}
\frac{(1-e^{-\tau})^2}{(1-e^{-\tau_*})^2}
\left(\frac{\tau_*}{\tau}\right)^{(0.2-\tau_*/3)}
\left(\frac{2}{l_{\rm pivot}}\right)^{n-1}
{A \over A_*} , \nonumber \\ 
{\cal C}_{Cl'} & = & {\cal C}_{C_* l}
\frac{(1-e^{-\tau})}{(1-e^{-\tau_*})}
\left(\frac{\tau_*}{\tau}\right)^{0.2}
\left(\frac{2}{l_{\rm pivot}}\right)^{n-1}
{A \over A_*} ,
\end{eqnarray}
where $l_{\rm pivot} \equiv (6000\ {\rm Mpc}\ k_{\rm pivot}) / 
\sqrt{\omega_m (1+z_{\rm ri})}$ 
and the initial potential power spectrum is taken
to be of the form  $k^3 P(k) = A\ (k/k_{\rm pivot})^{n-1}$ with 
$k_{\rm pivot}=0.05/{\rm Mpc}$.
The angular diameter distance shifting is achieved through the relation
$l' = l ({l'}_r/l_r)$ for ${\cal C}_{El}$ and 
$l' = l ({l'}_r+0.5)/(l_r+0.5)$ for ${\cal C}_{Cl}$.

The main features of the scaling are simple to elucidate. In the limit
of small $\tau$, the factor of $1-\exp(-\tau)$ gives the fraction of
scattered photons. The $\tau_\star/\tau$ factor is just a numerical fit
which makes Eq.~\ref{eq:pol} more accurate. 
The dependence on the temperature quadrupole to generate the
polarization anisotropy signal is clearly evident in the $n\neq 1$ pivot
factor. The projection formula in $l$ reflects the fact that most of the
action during reionization happens close to the actual epoch of
reionization (since the visibility function falls with increasing
time). The above approximate scalings are good to about 10\% around the
peak of the ``reionization bump'' for $\tau \la 0.3$. For our fiducial
(``$*$'') model we adopt a cosmology with $\tau=0.15$, $\omega_b=0.02$,
$\omega_m=0.16$ and $\Omega_\Lambda=0.65$. 

Although the process of reionization is expected to be spatially
inhomogeneous, this should have little effect on the large scale
polarization anisotropy discussed in the present study.  The 
details of the typical patch sizes of reionized ``bubbles'', and their
spatial correlations are not yet well--understood theoretically 
\citep{haardt99,ciardi00,gnedin00,miralda00,nakamoto01}.  However, it
appears likely that most of the ionizing radiation was provided by
galaxies, \citep[rather than quasars, for
example]{haiman02,ohhaiman02}, and the typical patch sizes are expected
to be ${\cal O}(1\ {\rm comoving \ Mpc})$. Inhomogeneities can remain
significant above the typical patch size scale because of spatial
correlations of the ionizing sources \citep{knox98,oh99}.  However, these
correlations fall off rapidly with increasing distance (approximately as
$1/r^{1.8}$). For the large angle polarization features described here
the relevant modes are $k\sim 2/\eta_r$; we do not expect correlations
to significantly affect such horizon scale features 
\citep[also see][]{gruzinov98,liu01}.

\section{Ionizing radiation and maximum optical depth}
\label{sec:taumax}

CMB anisotropy is sensitive to both cosmological parameters and
reionization history. As a result, any prior knowledge about
reionization can, in principle, tighten cosmological constraints, and
vice-versa.  For the most part, we will adopt a conservative
approach and obtain constraints without assuming any such {\em a
priori} knowledge.  In this section, however, we briefly consider
simplified semi--analytical models of reionization, and how they can
tighten constraints inferred from CMB data.  Reionization and its
dependence on cosmology has been studied more thoroughly by
\citet{venkatesan02} in the context of specific reionization
models. Here we restrict our analysis to obtaining a theoretical {\em
upper limit} to the reionization optical depth. In any cosmology with
Gaussian seed density fluctuations, the amount of non--linear mass
towards high redshifts declines exponentially, and hence a relatively
robust upper limit can be derived. 

In order to reionize the universe, the ionizing source must have (1)
produced at least $\sim1$ ionizing photon ($E>13.6$eV) per hydrogen
atom, and (2) the rate of production of ionizing photons must be
sufficient to balance recombinations, and keep the hydrogen ionized.
Assuming a uniform gas distribution at the mean IGM density, the
hydrogen recombination rate divided by the expansion rate is roughly
equal to $(11/(1+z))^{1.5}$, which implies that recombinations are
inevitably significant at $z\gsim 10$.  We therefore focus on the second
criterion above. 

The ionizing photon production rate in the early universe depends on
several factors: the rate at which gas is converted into new ionizing
sources; the initial mass function (IMF) of the ionizing source
population, the photon production rate per source (as a function of
mass), and the fraction of ionizing photons escaping into the IGM.  Here
we make the following conservative assumptions.  The rate of converting
gas into new ionizing sources is given by the fraction of mass collapsed
into dark matter halos with virial temperatures of $T>10^4$K (computed
using the formalism of \citet{press74}; see discussion and
justification in \citet{haiman02}).  We assume that all the gas
in these halos turns into ionizing sources, and produces $10^4$ ionizing
photons per baryon.  This is a conservative number; although massive
$\sim 1000~{\rm M_\odot}$ metal--free stars can produce upto $\sim 100$
times more ionizing photons \citep{tumlinson00, bromm01}, numerical 
simulations indicate that $\lsim 1\%$ of the available baryons
are expected to turn into ionizing sources \citep{abel00}.  We further
assume that all ionizing photons escape the source; in comparison,
escape fractions of $\sim 10\%$ are measured in local starburst galaxies
\citep{leitherer95} (although larger escape fractions have been
tentatively inferred for a sample of redshift $z\approx 3$ galaxies,
\cite{steidel01}).  Finally, in computing the global recombination rate,
we assume a mean clumping factor for ionized hydrogen of 
$C_{\rm int}\equiv\langle n_{\rm HII}^2\rangle /\bar n_{\rm HII}^2 =
20$; this is conservative in comparison with numerical simulations 
\citep{gnedin00,gnedin97}.

The above assumptions allow us to predict the maximum reionization
redshift,  and hence a maximum optical depth $\tau_{\rm max}$, given a
cosmology.  As an example, our formalism yields $z_{\rm max}=20.6$ and
$\tau_{\rm max}=0.18$ for a flat cosmology with $\omega_m=0.15$,
$\omega_b=0.02$ and $\sigma_8=1$.  Note that this upper  
limit is significantly higher than the expected ``ballpark'' values in
similar cosmologies \citep{haiman98,gnedin98}, $z\sim 7-10$.  We also
reiterate that our result for $\tau_{\rm max}$ is relatively insensitive
to our input assumptions, because of the steep decline in the collapsed
mass fraction towards high redshifts.  In our analysis of CMB anisotropy
data in the next section, we will study the effects of including the
constraint that the reionization optical depth not exceed 
$\tau_{\rm max}$. 

\section{Is the polarization feature detectable by MAP?}

In this section we examine how well MAP will be able to determine
$\tau$.  This question has already been answered for particular
models \citep{eisenstein99,tegmark00,prunet00,venkatesan02}.
However, the answer depends on the choice of cosmological
parameters, including $\tau$. Here we calculate the {\em probability
distribution} for $\sigma(\tau)$, the expected uncertainty
in the measurement of $\tau$, assuming flat adiabatic models of
structure formation as constrained by current CMB and quasar spectral
data.  

Calculating $P(\sigma(\tau))$ is a three--step process.  Step one is the
generation of a Markov chain of locations in the parameter space with the
remarkable property that the fraction of models in a given volume of
parameter space is equal to the posterior probability in that region;
i.e., a histogram of the frequency of occurence of parameter values in
the chain (``histogramming the chain'') gives the shape of the posterior
probability distribution function \citep{christensen01}. We assume a
flat cosmological model and take our (chain) parameters to be
$\omega_b$, $\omega_d$, $\Omega_\Lambda$, $A$, $n$, $x$ (where $\tau =
|x|$), four calibration parameters, and a parameter to account for the
Boomerang beam size uncertainty \citep{netterfield02}.  For this
section, we assume instantaneous reionization.  The data we use are from
DASI \citep{halverson02}, Boomerang \citep{netterfield02},
DMR \citep{bennett96} and Maxima \citep{lee01}.  Generation of the chain
requires as many likelihood evaluations as there are chain elements (in
this case 300,000 evaluations).  We use DASh \citep{kaplinghat02} which
rapidly and accurately calculates $C_l$ and performs the likelihood
calculation using the offset lognormal approximation of \citet{bond00}. 

Step two is the calculation, for each chain element, of ``derived
parameters'' (parameters that are derived from the chain parameters).
Our derived parameters include $z_{\rm ri}$, $\sigma_E(\tau)$ and
$\sigma_C(\tau)$.  The latter two are the errors expected on $\tau$ from
MAP's measurement of $C_l^E$ and $C_l^C$ respectively.  We calculate
$z_{\rm ri}$ by inverting Eq.~\ref{eq:taue}.
  
We calculate $\sigma_E(\tau)$ and $\sigma_C(\tau)$ by
\bea
\sigma_C^{-2}(\tau) &=& \sum_l 
\left({\partial C_{Cl} \over \partial \tau} / 
\Delta C_{Cl} \right)^2 \nonumber \\
\sigma_E^{-2}(\tau) &=& \sum_l 
\left({\partial C_{El} \over \partial \tau} / \Delta C_{El} \right)^2 
\eea
where (for low $l$) \citep{zaldarriaga97}
\bea
(\Delta C_{El})^2 &=&  2
{\left(C_{El} + w^{-1}\right)^2 \over (2l+1)f_{\rm sky}}\nonumber \\
(\Delta C_{Cl})^2 &=& 
{ C_{Cl}^2 + (C_{Tl} +w^{-1}/2)(C_{El} +w^{-1}) \over (2l+1)f_{\rm sky}}
\label{eq:clvariance}
\eea
and $w=2 \times 10^{14}$ is the expected weight--per--solid angle 
(for polarization measurements) from combining MAP's 90 GHz and 60 GHz
channels for two years. In calculating our results throughout the paper,
we set $f_{\rm sky}=1$ (since $f_{\rm sky}$ is expected to be close to
unity for the satellite experiments). Interested readers can scale our
results to any value of $f_{\rm sky}$ using Eq. \ref{eq:clvariance}. 
We calculate $C_{El}$ and $C_{Cl}$ at each point in the chain using
Eq. \ref{eq:pol}. The partial derivatives are calculated by finite
differencing and are accurate to $\sim 10\%$ around the peak of the
signal. Further, the derivatives are taken at fixed $A_\tau = A
\exp(-2\tau)$ since this quantity will be well--determined by high $l$
CMB measurements.  

Step three is the histogramming of the chain, with addition of various
prior constraints.  In this case, the priors are tophats implemented
by simply restricting the counting of models to those with $z_{\rm
ri}$ in the desired range.  

We have not included effects of foreground contamination because very
little is known about polarized emission from foregrounds in the
relevant frequency range. \citet{tegmark00} have tried to model our
ignorance of the foregrounds and use the CMB maps to learn more about
the foregrounds. The good news from their study is that foregrounds can
be removed, but this results in larger error bars for the estimated
cosmological parameters. With a ``conservative'' estimate 
for the amplitudes of foregrounds, \citet{tegmark00} showed that
the expected error in $\tau$ could increase by a factor of about 2
relative to the expected error assuming no foregrounds. For this
analysis they used all 5 MAP channels to fit the cosmological and
their foreground model parameters jointly. However, the amplitude 
of the polarized foregrounds is unknown. If this amplitude is a factor
of 10 larger (i.e., the power is a factor of 100 larger), then
\citet{tegmark00} estimate that the error on $\tau$ goes up by a 
factor of about 10.   

For simplicity, we assume MAP's 22, 30 and 40 GHz channels are
used entirely for foreground removal, allowing the total weight of the
60 and 94 GHz channels to be used for the CMB. Also, foregrounds may
significantly contaminate $C_{C l}$ but not $C_{E l}$, or vice--versa
and hence the existence of two different statistical channels to
determine $\tau$ increases our chances.  Note that the
errors $\sigma_E(\tau)$ and $\sigma_C(\tau)$ are correlated and this
must be taken into account if one were to combine them into a final
error for $\tau$.  

\begin{figure}[htbp]
  \begin{center}
    \plotone{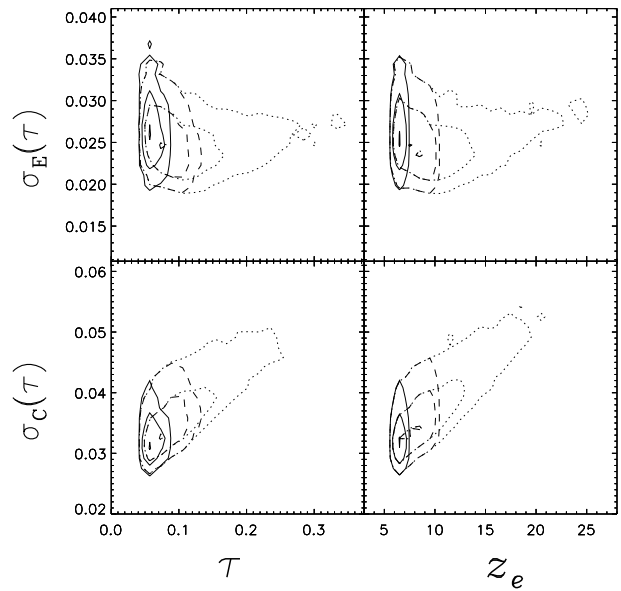}
    \caption{1 and 2-$\sigma$ joint probability contours of optical
    depth $\tau$ (or $z_e$) and the expected uncertainty with which MAP
    can measure $\tau$ in two years of operation. Measurements of $\tau$
    using the E (EE) and the C (TE) channels are considered
    separately. The left panels show the $\tau$, $\sigma_{E,C}(\tau)$
    contours and the right panels show the $z_e$, $\sigma_{E,C}(\tau)$
    contours. We have included three different priors on $z_e$. The
    dotted contours are for $z_e > 6$, the dashed contours are for
    $6<z_e<10$, and the solid contours are for $6<z_e<7$. } 
    \label{fig:mapsig}
  \end{center}
\end{figure}

We show the results of the exercise detailed in this section in
Fig.~\ref{fig:mapsig}. The four panels in the figure show contours of
equal probability (1 and 2-$\sigma$) in the $\sigma_E(\tau)$ and
$\sigma_C(\tau)$ vs. $\tau$ and $z_e$ planes. We have plotted these
contours for different priors on $z_e$. If reionization is indeed
instantaneous, then the $6 < z_e < 7$ prior (solid contours) is the
interesting case to study. However, as we stressed before, partial
reionization is possible at redshifts $z_e>6$, and hence we have
included contours with more conservative priors: $6 < z_e < 10$ (dashed
contours) and $6 < z_e$ (dotted contours). 

The result to take away from Figure \ref{fig:mapsig} is that MAP will be
able to determine $\tau$ to about $0.02$ to $0.03$ regardless of the
value of $\tau$. The lack of dependence of $\sigma_E(\tau)$ on $\tau$ is
due to two  
reasons. First, at its peak $C_{El} \propto \tau$ (that this is
approximately true can be ascertained from Eq. \ref{eq:pol}) and hence
around peak power $\partial C_{El}/\partial \tau $ is approximately 
constant. Second,  $\Delta C_{El}$ is dominated by the $w^{-1}$ noise
term rather than the $C_{El}$ sample variance term. The
cross-correlation power spectrum $C_{Cl}$ varies more slowly (than
linear) with $\tau$ and hence $\sigma_C(\tau)$ increases with $\tau$. 

Figure \ref{fig:mapsig} also includes the constraint that the optical
depth not exceed $\tau_{\rm max}$ calculated as outlined in Section
\ref{sec:taumax}. The effect of this $\tau_{\rm max}$ cut is to exclude
some of the large optical depth, $\tau \gsim 0.2$, models. Without
the  $\tau_{\rm max}$ cut, the 1-$\sigma$ contours extend to about 
$\tau=0.3$ for $\sigma_E(\tau)$ (and $\tau=0.25$ for $\sigma_C(\tau)$). 
With the cut, both the 1-$\sigma$ contours are restricted to
$\tau<0.2$. 
As mentioned in Section \ref{sec:taumax} the parameters we use in our 
semi-analytical model to calculate $\tau_{\rm max}$ are very
conservative. If the optical depth is in fact large, then reionization
models will provide significant complementary constraints. 

One can look at the detectability of the signal in another way. Let's
denote the value of the maximum $C_{E, l}$ for $l$ between 2 and 20 by
$C_{l, {\rm peak}}$. One can then include $C_{l, {\rm peak}}$ as a chain
parameter and obtain its probability distribution. A rough estimate of
the detectability of the signal can be obtained from looking at if 
$C_{l, {\rm peak}}$ (signal) is larger than the detector noise
contribution; for MAP this noise contribution is 
$w^{-1}=5\times 10^{-15}$.  

The distribution of $C_{l, {\rm peak}}$ peaks at 
$5.7\times 10^{-15}$. 90\% of the models with $z_e > 6.3$ in the chain
have  $C_{l, {\rm peak}} > 5.5 \times 10^{-15}$, which tells us that its
likely MAP will detect the reionization feature. If we now apply the
$\tau_{\rm max}$ cut, then  $C_{l, {\rm peak}} > 4.9 \times 10^{-15}$
for 90\% of the models. If we further restrict $z_e$ to between 6.3 and
7, then 90\% of the models have 
$C_{l, {\rm peak}} > 3.2 \times 10^{-15}$. 

We end this section by noting that since $A_\tau$ will be determined
with very high precision by MAP, the uncertainty in $A$ will be
dominated by the uncertainty in $\tau$.  If MAP detects the
reionization feature as discussed above, then the uncertainty in $A$
is expected to be  $\sigma(\ln{A}) = 2\sigma(\tau) = 0.04$ to  $0.06$.
Note that this is far better than what can be achieved using large
angle temperature anisotropy (COBE) data.    

\section{Is a gradual transition distinguishable from instantaneous?}

\begin{figure}[htbp]
  \begin{center}
    \plotone{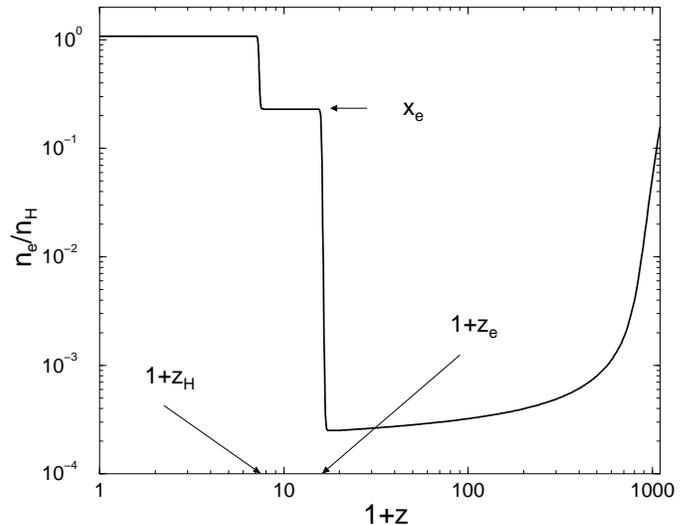}
    \caption{The evolution of the ionization fraction as a function of
 redshift (z) in a two-step reionization model. ${\rm n}_{\rm e}$ and
 ${\rm n}_{\rm H}$ are the number density of free electrons and hydrogen
 nuclei, respectively. The optical depth to the last scattering
 surface in this model is 0.05 and the epoch when electrons first
 appear is characterized by $z_e=15$ and $x_e=0.23$.}  
    \label{fig:2step}
  \end{center}
\end{figure}

The answer to this question depends on the details of how reionization
took place.  For simplicity, here we assume a two-step reionization
process (see Figure \ref{fig:2step} for the evolution of the 
ionization fraction in one such model). We assume that at some 
$z_{\rm  e}>z_{\rm H} = 6.3$, the universe was partially ionized, the
ionized fraction being $x_e \le 1.08$. The ionized fraction is defined
here as the ratio of number density of free electrons to that of
hydrogen nulei \footnote{ The present ionized fraction is slightly
  larger than unity because of the contribution from the complete
  reionization of $^4$He $\rightarrow$ $^4$He$^+$.}.  
At around $z=6.3$, new sources turned on, leading to almost complete
ionization. If reionization happened gradually (and not in our
idealized two--step fashion) then $x_e$ provides us with an estimate
of the average ionized fraction before $z=6.3$, and $z_e$ an estimate
of when the first ionizing sources turned on. As discussed above, the
large--scale polarization anisotropy should be insensitive to the
topology of the partially ionized regions.  

The power spectra used in the analyses described in this section were
calculated using the publicly available CMBfast code \citep{cmbfast}
which we modified to take into account the two-step reionization
process.  

We push the arguments outlined in Section \ref{sec:taumax}, 
to its limit, and assume a minimum virial temperature of
100 K for halos to cool \citep[via H$_2$ molecules; see][]{haiman00}
and form astrophysical objects that can
contribute to reionization.  Note that gas colder than 100 K
can not form H$_2$ and cool, and therefore can not contract or
fragment in the dark halo.   We find that this cooling cutoff
leads to the result: $z_e \lsim 40$. In order to have $z_e$
exceed this value, the presence of non-gaussianity in the
primordial power spectrum would be required.

\begin{figure}[htbp]
  \begin{center}
    \plotone{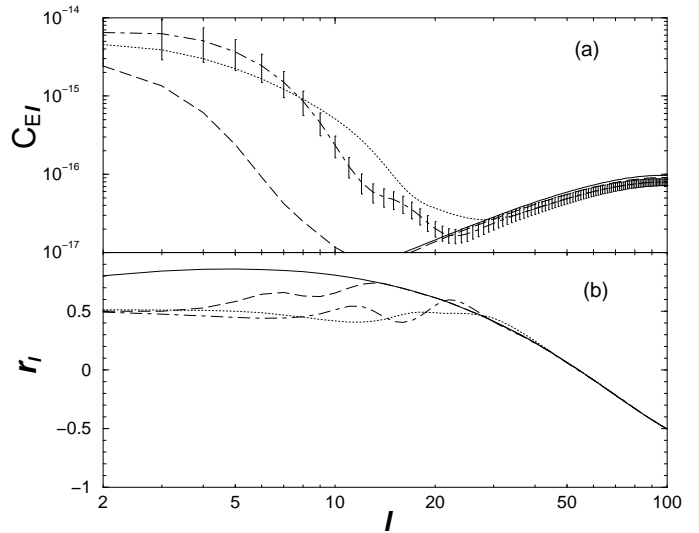}
    \caption{Effect of different reionization histories on the CMB
anisotropy. Panel (a) shows $C_{El}$, and (b) shows 
$r_l={\cal C}_{C l}/\sqrt{{\cal C}_{E l}{\cal C}_{Tl}}$. 
Both panels have the same set of reionization histories.
The solid curve shows the result of no reionization. 
The other 3 curves have $z_{\rm H}=6.3$. The dashed curve has $x_e=0$
which implies $\tau=0.033$ (note that $x_e$ is the ionized fraction for
$z>z_{\rm H}$). The dotted curve has $\tau=0.1$ and $z_e=30$ which
requires $x_e=0.26$. The dot-dashed curve has $\tau=0.1$ and $z_e=15$
which requires $x_e=0.89$. We have added cosmic variance error bars
to the $z_e=15$ $C_{E l}$ (dot-dashed) curve. } 
    \label{fig:spectra}
  \end{center}
\end{figure}

Figure \ref{fig:spectra} shows the effect of different reionization
histories. The polarization ``bump'' due to reionization is
unmistakable, as is the lack of power in the model with no
reionization (solid curve). The $\exp(-2 \tau)$ suppression at large $l$ 
is not very evident because of  the logarithmic scale for the E mode
polarization spectrum in panel (a). Note that the variable related to
cross-polarization plotted in panel (b) is the coefficient of
correlation $r_l=C_{C l}/\sqrt{C_{E l}C_{T l}}$
\citep{oliviera-costa02}.    

The curves to compare in Figure \ref{fig:spectra} are the dot-dashed and 
dotted ones. Both the curves have $\tau=0.1$; the dotted curve has 
$z_e = 15$ while the dot-dashed curve has $z_e=30$. The main effect of
increasing $z_e$ at fixed $\tau$ is to spread the total
reionization--induced power over a larger range of $l$. In fact, when
$z_e$ and $z_H$ are well separated, two reionization ``bumps'' can be
discerned. These are directly related to the bimodal nature of the
visibility function for our 2-step reionzation process. The change in
the shape of the polarization power spectra due to partial reionzation
at $z> z_H$ allows two models with the same optical depth but different
reionization histories to be distinguished from each other. For
reference we have plotted the cosmic variance error bars given by
$\sqrt{2/(2l+1)}C_{E,l}$ for the $z_e=15$ $C_{E,l}$ curve.

An interesting question, which is part of our broader goal
of distinguishing between models with differing ionized
fractions beyond $z=6.3$, is whether one can distinguish a completely 
neutral hydrogen content at $z>6.3$ from a partially ionized one.
We outline the method used to compare models with different
reionization histories below.

Given the spherical harmonic coefficients $a^X_{lm}$ derived from data,
one can write down the likelihood function for a particular model
($C_{Xl}$) given the data as:
\bea
-2 \ln({\cal L}) & = & \sum_{l} (2l+1)f_{\rm sky} 
\bigg[ \ln \left( [C_{El}+w^{-1}]\, C_{Tl}-C_{Cl}^2 \right) \nonumber \\ 
& + &
{[C_{El}+w^{-1}]\, C_{Tl}^d+C_{Tl}\, C_{El}^d-2C_{Cl}\, C_{Cl}^d \over
[C_{El}+w^{-1}]\, C_{Tl}-C_{Cl}^2} \bigg] 
\label{eq:like}
\eea
where $C^d_{Xl}\equiv \sum_m |a^X_{lm}|^2/(2l+1)$. Eq. \ref{eq:like}
was derived under the assumption that $a^X_{lm}$ are gaussian
distributed with mean zero. We have set the  
damping due to the finite experimental beam width to be unity (since we
are only interested in the low $l$ region), and omitted the noise term
for $C_{Tl}$ (since it's negligible in comparison to $C_{Tl}$). 
We have no data in hand. Instead, we will assume that $C_{Xl}^d$ is
given by some fiducial model, $C_{Xl}^{\rm fid}$ (in other words we
perform an ensemble average over $\ln({\cal L})$), such that
$C_{Xl}^d=C_{Xl}^{\rm fid}+w_X^{-1}$ with $w_C^{-1}=0$. 
We then look at the likelihood of other models in this parameter
space. Note that all the discriminatory power comes from $l \la 30$. In
this regime (as we have shown in Eq. \ref{eq:pol}) the only
cosmological parameters the spectra are sensitive to are the amplitude
of power spectrum, tilt and the optical depth. 

\begin{figure}[tbh]
\plotone{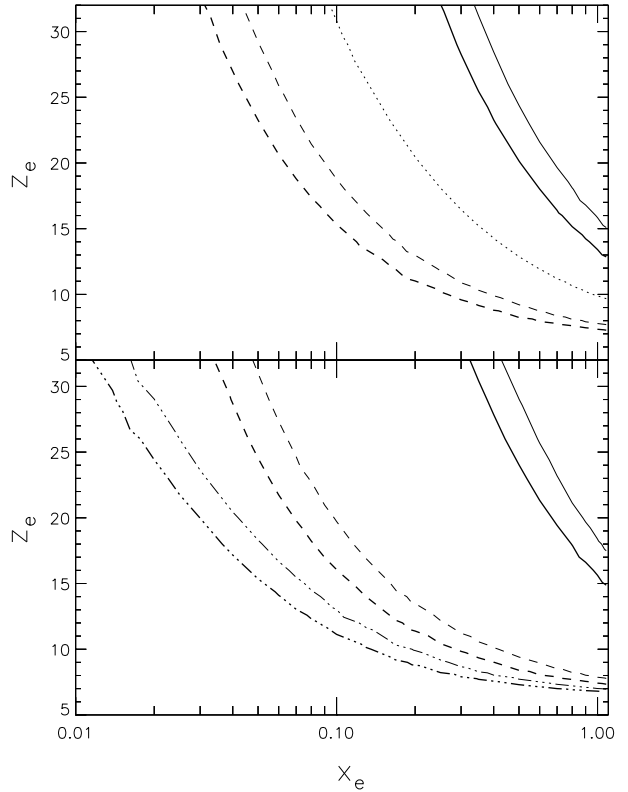}
\caption{Constant likelihood contours at 10\% (thick curves) and 1\%
(thin curves) of the maximum likelihood. The maximum likelihood occurs
at the fiducial model, which for this plot is one with no ionized
hydrogen at $z>6.3$. The upper panel assumes two MAP
and Planck channels as described in the text. The lower panel assumes 
one channel per experiment. The solid curves are for MAP. The dashed
curves are for Planck. The dot-dashed curves are for a hypothetical
SuperPlanck  (described in the text). The dotted curve is a constant 
$\tau=0.06$ contour. } 
\label{fig:zxabrupt}
\end{figure}

\begin{figure}[tbh]
\plotone{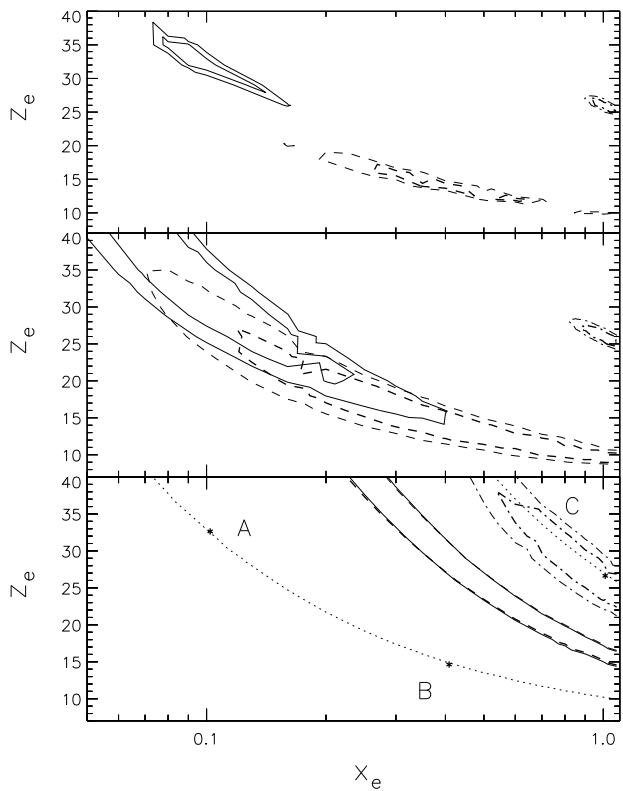}
\caption{Constant likelihood contours at 10\% (thick curves) and 1\%
(thin curves) of the maximum likelihood -- which occurs at the fiducial
models labeled by A, B and C and denoted by asterisks. 
A has $z_e=32$ and $x_e=0.1$; 
B has $z_e=14$ and $x_e=0.4$; 
C has $z_e=26$ and $x_e=1.0$. 
All the panels have the same fiducial models. The solid curves
correspond to model A, the dashed curves to model B and the dot-dashed
curves to model C. 
The upper panel is for a cosmic variance limited experiment,
``SuperPlanck'' (see text for details). 
The middle panel assumes 1 Planck channel sensitivity while the lower
panel is for two channel MAP sensitivity. The dotted curves in the lower
panel are curves of constant $\tau$ for $\tau=0.063$ and $\tau=0.25$.} 
\label{fig:zx}
\end{figure}

To fix the amplitude we again use the fact that the CMB constrains the
combination $A_\tau=A\exp(-2\tau)$ very well 
-- to about 10\% (at 1-$\sigma$) with present data (without including
massive neutrinos). This will greatly improve with MAP. Note that this
particular combination of the amplitude of the power spectrum and
optical depth is not affected by the tensor contribution to the
anisotropy since the constraints come from the small angular scales. The
error in $A_\tau$ when propagated to the $C_l$s is much smaller than
cosmic variance on the scales that we consider ($l<30$). This is simple
to see: since the $C_l$s scale linearly with $A$, for any given $\tau$
the fractional theoretical error in $C_{Xl}$ due to the uncertainty in
$A_\tau$ is $\sigma(A_\tau)/A_\tau$. Thus even with current data this
fractional error becomes larger than cosmic variance only at $l>100$ --
safely outside the regime we are interested in. 

We set the tilt to unity. Future CMB data sets will determine the tilt
well -- the expected errors are $\sigma(n) = 0.05$ for MAP and
$\sigma(n) = 0.02$ for Planck (allowing for small variations of the
spectral index with scale) \citep{eisenstein99}. Relaxing this
constraint will not affect our results. For $C_{El}$ and $C_{Cl}$, the
predominant effect of tilt (at small $l$) is just a rescaling of the
spectra -- degenerate with the  amplitude of the power spectrum. The
fractional change in $C_{C,E l}$ due to a non-zero $n-1$ is $\simeq
4.5(n-1)$. Given the $l$-range of the signal and the precision with
which $n$ will be determined, this is not an issue. 

The experiments we consider here are MAP and Planck. To be conservative,
as mentioned, we only assume a maximum of two MAP or Planck channels. The
best guess at the two relatively foreground--free Planck channels are the
143 and 100 GHz channels \citep{tegmark00}. The weight--per--solid angle
(for polarization measurements) of the 143 GHz channel for one year of
observation is expected to be $1.35 \times 10^{16}$ and from combining  
the 100 and 143 GHz channels it's expected that $w = 1.67 \times
10^{16}$. 

We will not consider here the balloon borne and terrestrial polarization
experiments since these experiments will not have the required sky
coverage to obtain spectra at the large angular scales
at which reionization imprints its signal. 

Figure \ref{fig:zxabrupt} shows constant likelihood contours for the
case where the fiducial model has only neutral hydrogen
at $z>6.3$. The constant likelihood contours are drawn at 10\% and 1\%
of the maximum  likelihood. We have considered two separate cases
plotted in the upper and lower panels  corresponding to sensitivities
equal to that of MAP and Planck assuming two channels or a single
channel. This is important for MAP since each channel has equal weight.  

In the lower panel of Figure \ref{fig:zxabrupt} we also plot the results
for an experiment we dub ``SuperPlanck'' 
with a sensitivity which is 25 times better than Planck, i.e., 
$w=3.4\times 10^{17}$. For reference, the raw sensitivity ($1/\sqrt{w}$)
of Planck is a factor of 11.6 better than MAP. Clearly, the results
of using SuperPlanck are remarkable. We have chosen SuperPlanck to be 25
times more sensitive than Planck because that makes SuperPlanck nearly
cosmic variance limited for $z_e < 30$. The sensitivity of
SuperPlanck also makes it a great tool to hunt for primordial (tensor)
B-mode polarization signal. This signal would be detectable by
SuperPlanck if the ratio of tensor to scalar pertuburbations is larger
than 0.001 \citep{knox02,kesden02} which implies an energy scale of 
inflation greater than $6 \times 10^{15} {\rm GeV}$.  We study the 
results of Figure \ref{fig:zxabrupt} in more detail below. 

Let us first concentrate on MAP. Using the 10\% likelihood contours, we
see that MAP should be able to distinguish a completely neutral
hydrogen content at $z>6.3$ from a case where hydrogen is 50\% ionized
between the redshifts of $z_{\rm H}=6.3$ and $z_e=20$. We see that in
this region of parameter space the contours are essentially parallel to
the constant $\tau$ contours -- implying there is very little
information except for the value of $\tau$ itself. 

Planck, of course, can do a much better job. Planck can distinguish
between 10\% ionized fraction between $z_{\rm H}$ and $z_e=15$ and the
fiducial model which has $x_e = 0$ at $z > z_{\rm H}=6.3$. The
hypothetical SuperPlanck could distinguish the fiducial
abrupt--reionization model from one in which $x_e=0.05$ between
$z_{\rm H}$ and $z_e=15$. 

We note that if the universe was indeed reionized close to
$z=6.3$, then MAP will be able to show that $z_e < 13, \ 15$ at 
2, 3-$\sigma$. Clearly, this will not be sufficient to show that
the universe was really abruptly reionized close to $z=6.3$. However,
for the same assumptions, the upper bound from Planck will be 
$z_e<7.5,\ 8$. This will be a strong indication that most of the
reionization took place close to $z=6.3$, although delays of 
$\Delta z \simeq 1$ will still remain a possibility.
These results can be read off from Figure \ref{fig:zxabrupt} (the 10\%
and 1\% contours projected to the $x_e=1$ line aproximately give the 2
and 3 $\sigma$ bounds), or derived from Figure \ref{fig:mapsig}. From
Figure \ref{fig:mapsig} we see that $\sigma(\tau) \simeq 0.025$ (for
MAP) which implies  $\sigma(z_e) \simeq (2/3) z_e \sigma(\tau)/\tau = 3$
for $z_e=6.3$. If we assume that both MAP and Planck are noise (and not
cosmic variance) limited for $z_e=6.3$, then we have 
$\sigma(z_e) \simeq 0.4$ for Planck. This is somewhat better than what
Figure \ref{fig:zxabrupt} implies because cosmic variance is not
completely negligible.

One could also ask the following question: if
the real universe was completely reionized at some $z>6.3$, then at 
what $z$ does it become possible to be certain (say at 3-$\sigma$) that
$z_e>6.3$? We can use the discussion of the previous paragraph to answer
this question. Taking $\tau \simeq 0.002 z^{1.5}$, we have that
$\sigma(z_e) \simeq 333\sigma(\tau)/\sqrt{z_e}$ and to a good
approximation we can assume that $\sigma(\tau)$ is independent of $\tau$
for the optical depths we are considering here. Thus if the universe
was reionized at a redshift of 13 (8), then MAP (Planck) will be able to
claim that $z_e > 6.3$ at the 3-$\sigma$ level.

We take the above discussion to its natural conclusion by answering the
following question: what information can one obtain if the universe is
partially ionized at $z > 6.3$?  
Figure \ref{fig:zx} is similar to Figure \ref{fig:zxabrupt} except we
now plot likelihood contours for different fiducial models. Specifically
we consider 3 fiducial models: (A) $z_e=32$ and $x_e=0.1$, 
(B) $z_e=14$ and $x_e=0.4$, (C) $z_e=26$ and $x_e=1.0$. 
Models (A) and (B) have the same optical depth of $\tau = 0.065$, while
model (C) has a large optical depth of $\tau=0.25$. We see that at the
larger optical depth, MAP is sensitive to more than just $\tau$. The
distribution of power in $l$ which is decided by $z_e$ is sufficiently
different that the contour does not follow the constant $\tau$ contour
along the entire parameter space. For Planck this happens at the much
lower $\tau$  of $0.065$. It is clear from the Planck contours (upper
panel) that cases (A) and (B) can be distinguished despite their equal
optical depths.  

This is an interesting result. It opens up the possibility that we can
observationally determine the reionization history of the universe. The
usefulness of such data will be directly determined by the
sophistication of the physical modeling of the end dark ages is. We end
this section with the tantalizing suggestion that with SuperPlanck, it
should be possible to map the evolution of the ionized hydrogen fraction 
$x_e(z)$ by breaking up the $z_{\rm H}$--$z_e$ redshift interval into
several bins and tracing the evolution of $x_e$. 

\section{Conclusions}

The remarkable discovery of Gunn-Peterson trough in the spectrum of the
redshift $z=6.28$ quasar has brought into sharp focus the questions of
how and when the universe was reionized. While representing a
significant new discovery, the power of the hydrogen Gunn--Peterson
trough to probe the dark age is limited, because a small neutral
hydrogen fraction ($x_{\rm H} \sim 10^{-3}$) is enough to give rise to
such a trough. CMB polarization is the most promising way to probe
deeper into the dark age; our goal in this paper has been to quantify
this statement for forthcoming CMB anisotropy experiments. 

Contaminating astrophysical (foreground) sources of polarization 
may be present at significant levels, but there is very little
we know about them.  In deriving our results, we have used a maximum of
two channels per experiment, with the remainder assumed to be used as
foreground monitors.  Foregrounds might be more detrimental than we have
assumed; the only way to find out is to do multi--frequency CMB
experiments. 

 
Using current data we have shown that MAP will most likely detect the
reionization feature in the polarization data. We have shown that MAP
can measure the electron optical depth to 0.02-0.03 (1-$\sigma$),
regardless of the value of the optical depth and that this measurement is
sufficient to break the degeneracy between the optical depth and the
amplitude of primordial density perturbations. Further, MAP (with two
years of observation) will be able to distinguish a model with only
neutral hydrogen at $z>6.3$ from one that was about 50\% ionized between
$z=6.3$ and 20. 
Planck, with its much higher projected sensitivity (and one year of 
observation), will be able to distinguish an ionized hydrogen fraction
of about 6\% between the redshifts of 6.3 and 20 from the model with no
ionized hydrogen at $z>6.3$. More strikingly, Planck (and MAP if the
optical depth is larger than about 0.2) should be able to differentiate
among reionization histories which lead to the {\em same} optical depth
for CMB photons.  These results will enormously deepen our understanding
of the physics of reionization.

\acknowledgements

ZH was supported by NASA through the Hubble Fellowship grant
HF-01119.01-99A, awarded by the Space Telescope Science Institute, which
is operated by the Association of Universities for Research in
Astronomy, Inc., for NASA under contract NAS 5-26555. LK and MK were
supported by NASA grant NAG5-11098. GPH was supported by W.M. Keck
foundation. We thank Uros Seljak and Matias Zaldarriaga for the use of
their CMBFAST code. 

\bibliography{reion}

\end{document}